# Photoluminescence of InGaAs/GaAsBi/InGaAs type-II quantum well grown by gas source molecular beam epitaxy

Wenwu Pan[1,2], Liang Zhu[2,3], Liyao Zhang[1], Yaoyao Li[1], Peng Wang[1,2], Xiaoyan Wu[1,2], Fan Zhang[4], Jun Shao[3], Shumin Wang[1,5]

[1]State Key Laboratory of Functional Materials for Informatics, Shanghai Institute of Microsystem and Information Technology, CAS, 865 Changning Road, Shanghai 200050, China

[2]University of Chinese Academy of Sciences, Chinese Academy of Sciences, Beijing 100190, China

[3]National Laboratory for Infrared Physics, Chinese Academy of Sciences, 500 Yutian Road, Shanghai, 200083, China

[4]School of Physical Science and Technology, ShanghaiTech University, Shanghai 201210, China

[5]Department of Microtechnology and Nanoscience, Chalmers University of Technology, Gothenburg 41296, Sweden

Email: shumin@mail.sim.ac.cn, lyzhang@mail.sim.ac.cn

## Abstract

$In_{0.2}Ga_{0.8}As/GaAs_{1-y}Bi_y/In_{0.2}Ga_{0.8}As$ ($y \geq 3.4\%$) quantum wells (QWs) were grown on GaAs substrates by gas source molecular beam epitaxy for realizing the type II band-edge line-up. Both type I and type II transitions were observed in the Bi containing W QWs and the photoluminescence intensity was enhanced in the sample with a high Bi content, which is mainly due to the improvement of carrier confinement. Blue-shift of type II transitions at high excitation power density was observed and ascribed to the band-bending effect. The calculated transition energies based on 8 band ***k·p*** model fit well with the experiment results. The experimental and theoretical results show that the type-II QW design is a new promising candidate for realizing long wavelength GaAs-based light emitting devices near 1.3 μm.

## 1. Introduction

Recently, GaAsBi compounds have been a subject of intensive theoretical and experimental studies due to their potential applications in laser diodes with effective extension of light emitting wavelength, a relatively temperature-insensitive wavelength and reduction of Auger recombination [1-5]. The incorporation of a small amount of Bi atoms into GaAs strongly reduces band-gap energy (84-91 meV/Bi%) [1, 2, 6] and, in addition, decreases the temperature coefficient of the band gap from 0.45 meV/K to about 0.15 meV/K [3]. In spite of the success of growing GaAsBi with Bi content up to 22% [7] and the evident progress that electrically pumped GaAsBi quantum well (QW) laser has been fabricated successfully with the lasing wavelength up to 1.06 μm [8], the room temperature emission at longer wavelengths is rather limited. This is due primarily to the difficulty of incorporating a high Bi alloy fraction and simultaneously retain high optical quality. The growth parameter most sensitive to the optical quality as well as the Bi content is growth temperature. For example, with an increase of the temperature from 350 °C to 380 °C, the Bi content dramatically decreased from 4.5% to 0.6% due to weak Ga-Bi bond and the strong tendency for Bi to surface segregate in GaAsBi [9]. On the other hand, the Bi concentration increased from 1% to 5% while photoluminescence intensity (PL) was observed to drop by more than a factor of 1000 when the growth temperature was decreased from 400 °C to 300 °C with all other growth conditions kept fixed [10]. The reported maximum growth temperature for fabricating device is about 380 °C and about 4% Bi is incorporated in the GaAsBi active layer [11]. Thus emission beyond 1.3 μm is difficult to be realized by using conventional type-I GaAsBi QWs grown on GaAs.

Type-II heterostructures employing the InGaAs(N)/GaAsSb [12, 13] and InGaAs/GaAsN [14, 15] system on GaAs have been previously demonstrated to be promising new materials for laser emission at 1.55 μm. Problem associated with the N incorporation is that the optical quality degrades significantly with the N content increasing [16]. For bismide compounds, previous studies are focused on type I heterojunctions like (In)GaAsBi/(Al)GaAs [17, 18] and to our knowledge no experiment work related to type II heterostructures have been reported. Only a theoretical work on the type II GaAsBi/GaAsN system based on 8 bands **k·p** method have been reported, suggesting it may provide a new material system with lattice match to GaAs in a spectral range of significant importance for optoelectronic devices [19].

The band offset of GaAs/GaAsBi heterojunction is type I and the band offset ratio in conduction and valence band are 29/91 and 62/91, respectively [6, 20]. The large valence band offset (VBO) ratio leading to good confinement of holes in GaAsBi. The conduction band offset (CBO) ratio in strained $In_yGa_{1-y}As$/GaAs QW ($0.18<y<0.25$) is 0.6 [21] which is much larger than that of strained GaAsBi/GaAs QW. Similar to InGaAs/GaAsSb, type II band alignment can be expected in $In_yGa_{1-y}As$/$GaAs_{1-x}Bi_x$ heterojunction when the In and Bi content satisfying the relationship of $3<y/x<15$. In addition, the reduction of band gap is about 0.26 eV per 1% of compressive strain in GaAsBi, which is almost 3.5 times as large as Sb-related reduction of band gap in GaAsSb alloy grown on GaAs [22, 23]. It is implied that the type II GaAsBi/InGaAs QW can be a promising new structure for GaAs-based laser emission beyond 1.3 μm and we are therefore very interested to predict its effective band gap energy using the **k·p** method.

In this study, InGaAs/GaAsBi/InGaAs heterostructures were grown by gas source molecular beam epitaxy (GSMBE) and both type I and type II transitions are demonstrated at low temperature (LT). The experimental and theoretical results based on an 8 band **k·p** model shows that the InGaAs/GaAsBi W heterostructure QW has potentials to extend the light emitting wavelength to 1.3 μm.

## 2. Experiments

Samples were grown on semi-insulating GaAs (001) substrates using a V90 GSMBE system equipped with effusion cells for Ga and Bi. $As_2$ was cracked from arsine at 1000 ℃. A sandwich structure consists of 8.5 nm $In_{0.2}Ga_{0.8}As$, 7.8 nm GaAs(Bi) and 8.5 nm $In_{0.2}Ga_{0.8}As$ layers, as shown in the inset of Fig. 1. (a). The substrate temperature was measured by a thermocouple. After oxide desorption performed at 635 ℃, a 100 nm thick GaAs buffer layer was grown. The growth rate of GaAs and the pressure of $AH_3$ were fixed at 845 nm/h and 650 Torr, respectively. For the growth of InGaAs/GaAs(Bi)/InGaAs active region, the substrate temperature was decreased to 410 ℃ which is close to the typical growth temperature of 380 ℃ reported in Ref. 11. The $AH_3$ pressure was decreased to 300 Torr when growing the GaAsBi layer to facilitate Bi incorporation. The growth was finished by deposit of a 100 nm thick GaAs cap layer at 635 ℃ and the *in-situ* annealing effect to improve material quality is expected [24]. Two samples (labeled as S2, S3) with different Bi contents in the GaAsBi layer were grown, respectively. As references, an InGaAs/GaAs/InGaAs QW sample (labeled as S1) and two single 7.8 nm GaAsBi QW samples (labeled as S4, S5) were also grown with the same In and Bi composition as in the InGaAs/GaAsBi/InGaAs QW, respectively.

Structural properties of the InGaAs/GaAs(Bi)/InGaAs QWs were investigated by the measurements of (004) ω/2θ scans using a Philips X'Pert Epitaxy high resolution X-ray diffractometer (HRXRD) equipped with a four-crystal Ge (220) monochromator. Cross-sectional transmission electron microscopy (TEM) images and energy dispersive X-ray spectroscopy (EDX) were taken in order to investigate the crystalline quality, the homogeneity of the Bi composition as well as layer thickness in more detail. The PL spectra were measured using a Nicolet Megna 860 Fourier transform infrared (FTIR) spectrometer, in which an InGaAs detector and a $CaF_2$ beam splitter were used. A diode pumped solid-state (DPSS) laser (λ =639 nm) was used as the excitation source. 8 band *k·p* model was employed to calculate transition energy of the type II InGaAs/GaAsBi/InGaAs QWs.

## 3. Results and Discussions

The HRXRD (004) ω/2θ curves are shown in Fig. 1. (a). In each curve, the relatively narrow and strong peak comes from the GaAs substrate as well as GaAs epitaxial layers, while the broad and weak peak corresponds to the InGaAs/GaAs(Bi)/InGaAs QW. It is obvious that the signal of the InGaAs/GaAsBi QW shifts to a lower angle with respect to that of the InGaAs/GaAs QW labeled S1, indicating a higher average lattice constant of the InGaAs/GaAsBi QW. Satellite peaks between 31.7° and 32.7° can be seen in the measured HRXRD curves for the three samples, indicating that the quality of both the QW and the interface is very good. Figure 1. (a) also shows the simulated curves of S1 by setting In content to 20% in two 8.5 nm InGaAs layers which are separated by 7.8 nm GaAs layer. For S2 and S3, it should be noted that the determination of Bi content in QW are sensitive not only to the X-ray data obtained but also to the assumed shape of the concentration profile. Figure 1. (b) shows a high-resolution cross section TEM image of S3 and a typical EDX line scan profile measured along the growth direction. As shown in the figure, no dislocations is observed indicating the layers remain stain after the *in-situ* annealing at 635 ℃. The In composition and the thickness of each layer is very close to the original design of reference sample S1. Note that, the Bi content cannot be directly inferred from the EDX profile as the QW is too thin for EDX measurement to provide a direct correspondence with the absolute Bi content. By analyzing intensity profiles taken across the W QW region on the TEM image, it is shown that Bi segregation in the growth direction takes place, leading to an asymmetric Bi concentration profile in InGaAs/GaAsBi/InGaAs QW. The highest concentration of Bi was found much closer to the upper GaAsBi/InGaAs interface. This makes the InGaAs/GaAsBi very rough whereas the GaAsBi/InGaAs interface is much sharper. The InGaAs/GaAsBi interface fluctuates over about three monolayers. Thus unexpected InGaAsBi alloy may be formed near the InGaAs/GaAsBi interface. The Bi segregation is caused by the relatively high growth temperature for GaAsBi. The mean In content in the two InGaAs layers of S2 and S3 were estimated to be the same with S1, i.e., 20%, while the average Bi content in S2 and S3 are deduced from PL of two single GaAsBi QW samples grown under the same growth conditions, labeled S4 and S5, respectively, based on comprehensive analysis of their PL spectra and theoretical calculation results which will be shown below.

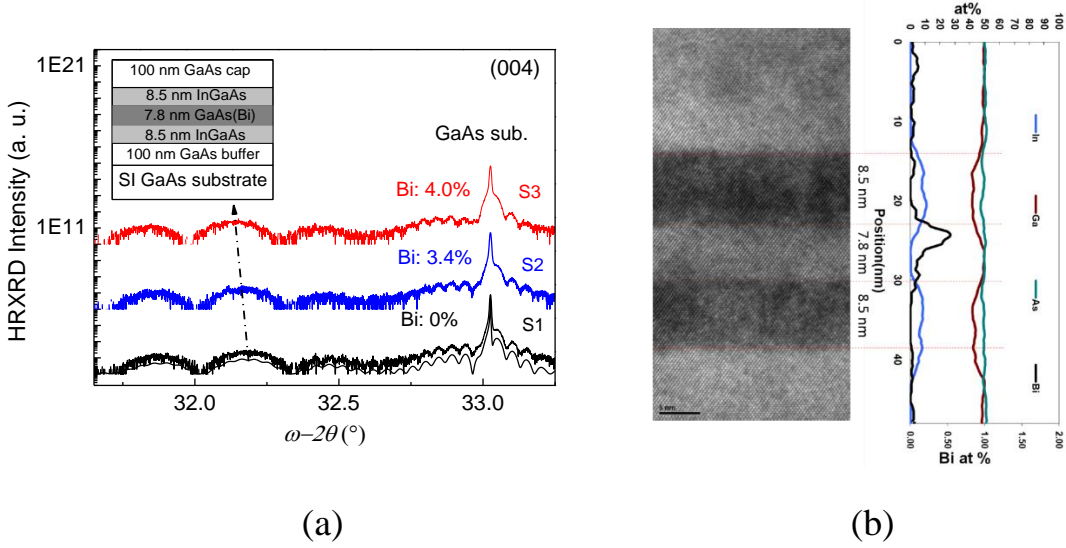

(a)            (b)

Figure 1. (a) HRXRD (004) $\omega/2\theta$ curves of the InGaAs/GaAs(Bi)/InGaAs W QW structures (S1、S2、S3) with various Bi contents in the GaAsBi. The solid lines are the simulated curves of samples by setting In content to 20% in two 8.5 nm InGaAs layers separated by 7.8 nm GaAs layer. (b). Typical TEM image with high-resolution modes of a cross section of S3 and typical EDX line scan profile measured along the growth direction.

Figure 2 shows the 8 K PL spectra of S2 and S3 using 80 mW excitation. As reference, the PL spectra measured under the same condition of S1, S4 and S5, are also shown. For S4 and S5, a single PL peak is observed at 1.284 eV and 1.238 eV with a full width at half maximum (FWHM) of 114 meV and 138 meV, respectively, which corresponds to the calculated type I transitions of GaAsBi QW with a Bi content of 3.4% and 4.0%, respectively, assuming 91 meV/Bi% band gap bowing effect as well as full strain in the QW. For InGaAs/GaAs(Bi)/InGaAs QWs, S2 and S3 show two peaks between 1.1 eV to 1.4 eV while no obvious PL is observed in the S1. The peaks with a similar intensity at 1.302 eV (P21) and 1.315 eV (P31) corresponds to the transition of localized electrons in conduction band and delocalized holes in valence band of its $In_{0.2}Ga_{0.8}As$ layers, respectively, as seen in the inset of Fig. 2. The slight energy difference of 13 meV suggests about 1% In composition difference between S2 and S3. Under the same excitation, no obvious PL features of InGaAs are observed in S1, indicating poor crystalline quality of the InGaAs QWs grown at this low temperature of 410 ℃. However, after Bi incorporation, confinements of both electrons and holes are enhanced due to the band gap bowing in the middle GaAsBi layer. Also, it is expected that Bi acts as surfactant [25] to promote interface smoothness and improve PL efficiency in GaAs and InGaAs layer. Similar effects have been observed in GaAsBi/AlGaAs QWs [18]. The FWHMs of P21 and P31 is about 25 meV which is larger than that of 10-15 meV from a typical single InGaAs/GaAs QW [26] and can be ascribed to the roughness of the InGaAs/GaAsBi interface or poor crystalline quality due to low temperature growth of InGaAs.

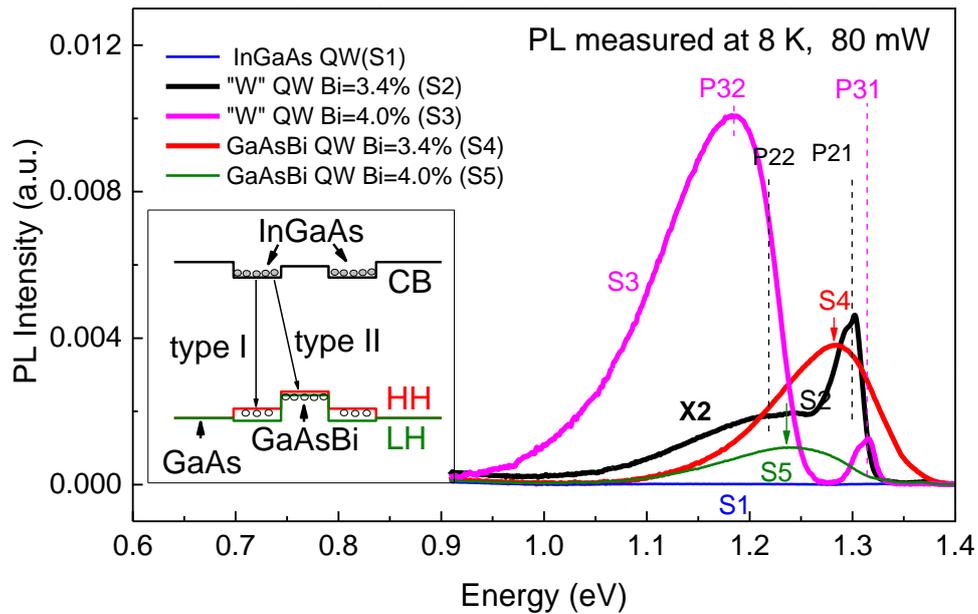

Figure 2. PL spectra of the InGaAs/GaAs(Bi)/InGaAs W QW structures (S1、S2、S3) and two GaAsBi/GaAs single QWs (S4、S5) with excitation of 80 mW at 8 K. The transition type I and II are illustrated by an inset in the figure.

The two broaden peaks at low energy of 1.210 eV (P22) and 1.184 (P32) eV are observed in S2 and S3, respectively. Since the energy is lower than the transition energy of the respective InGaAs QW and the GaAsBi QW, we attribute both to the type II transition between electrons in the InGaAs conduction band and holes confined in the GaAsBi valence band as shown in the inset of Fig. 2. For S2 with 3.4% Bi content, the P21 originated from recombination of the InGaAs layers dominates and the observed relatively weak shoulder contributed by P22 suggests that the VBO of GaAsBi$_{0.034}$/In$_{0.2}$Ga$_{0.8}$As is too shallow to localize most of holes with excitation of 80 mW at 8 K. When increasing the Bi content to 4% in S3, the VBO of GaAsBi/InGaAs increases and most holes are confined in the GaAsBi layer causing the type II radiative recombination dominating. Also, the Bi segregation shown in Fig. 1. (b) will lead a rich-Bi region near the InGaAs/GaAsBi interface and produce an approximately triangular well, which will enhance the overlap of the electron wavefunction in the upper InGaAs and hole wavefunction in the GaAsBi layer, leading to the enhancement of type II PL. The VBO value of the In$_{0.2}$Ga$_{0.8}$As/GaAsBi$_{0.04}$ heterojunction is estimated to be 169 meV which is close to the sum of the energy difference (P31-P32), i.e., 131 meV and the blueshift of 26 meV due to the band bending effect discussed below. The FWHM of P22 and P32 are about 190 meV and 138 meV, respectively, much broader than that of P21 and P31. Such a broadening effect is related to the non-uniform distribution of Bi and/or Bi clusters in the GaAsBi layer [27, 28].

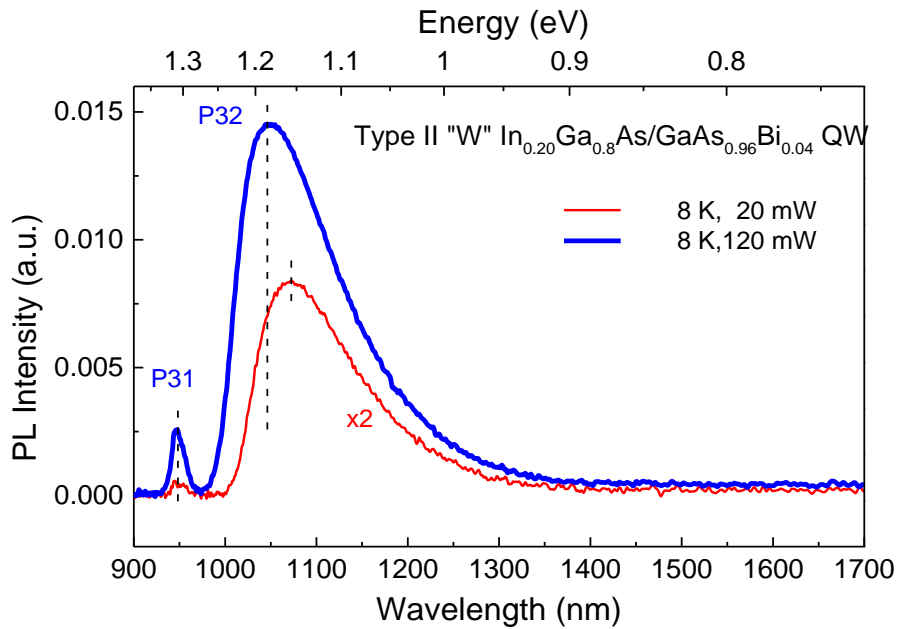

Figure 3. Low temperature (8 K) PL of the Type-II W In$_{0.2}$Ga$_{0.8}$As/GaAsBi$_{0.04}$ QW (S3) excited at 20 and 120 mW, respectively.

Figure 3 shows low temperature (8 K) PL spectra of Type II W QW S3 with excitation of 20 and 120 mW, respectively. With the exciting power increasing from 20 mW to 120 mW, the wavelength shifts from 1063 nm (1.166 eV) with an FWHM of 132 meV to about 1040 nm (1.192 eV) with a larger FWHM of 142 meV at 8 K. The 26 meV blue-shift at a higher excitation power density is due to the band-bending effect caused by spatial separation of electrons in the InGaAs QW and holes in the GaAsBi QW, which becomes strong. The band bending effect has also been seen in the type II luminescence from SiGe QWs [29] and InGaAs/GaAsSb heterojunctions [12]. Also shown in Fig. 3, the near band gap type I transition energy of P31 from the InGaAs layer is virtually not affected by the change in the excitation power, which strongly supports the type II assignment to the transition P32 in Fig. 2 and Fig. 3.

Table 1. III-V parameters and nonzero bowing parameters used in 8 band $\mathbf{k \cdot p}$ model for calculating transition energies in (In)GaAs/GaAs(Bi)/(In)GaAs QWs.

| Parameters | InAs | GaAs | GaBi |
|---|---|---|---|
| Band gap | 0.417 | 1.519 | -1.45 |
| Valence band position | -0.47 | -0.8 | -0.2 |
| Spin-orbit split-off energy | 0.39 | 0.341 | 2.2 |
| Bowing parameters | Band gap | Valence band offset | |
| $In_xGa_{1-x}As$ | 0.477 | 0.13 | |
| $GaAs_{1-x}Bi_x$ | 6.131 | -5.6 | |

To aid in interpreting the PL experiments and provide a general tool for predicting the effective band-gap energy for this kind of QW, i.e., the type II transition energy, 8 band $\mathbf{k \cdot p}$ model, which was demonstrated to be effective to calculate material gain in dilute bismides QW [23], was used. Previous studies show that the band structure was primarily dependent on the structure design, strain parameters, the band gap energy and band offsets of the materials [19]. The Bi incorporation significantly pushes up the VB edge. For ternary GaAsBi alloy, the Bi-related changes in the conduction and valence bands are taken from recent *ab-initio* calculations to be 29 meV/Bi% and 62 meV/Bi%, respectively [6, 20]. Table 1 shows the band parameters of InAs, GaAs and GaBi and the nonzero bowing constant for InGaAs and GaAsBi which have a high degree of certainty for our calculations [2, 30]. The calculations were performed by using Heterostructure Design Studio software [31]. The variation of the effective band gap with In and Bi content was shown in Fig. 4. The calculated results shows that about 6.5% Bi is needed to achieve emitting at 1.3 μm for the traditional GaAs/GaAsBi QW at room termperature. However it should be noted that the relatively high Bi content is hard to achieve with high optical quality and no laser device so far has been fabricated successfully with Bi content beyond 6% yet. Based on the type II InGaAs/GaAsBi QW, emission at 1.3 μm can be realized with about 40% In and 4.5% Bi incorporated in the InGaAs and GaAsBi layer, respectively, which is realistic for MBE growth to achieve good optical quality at present.

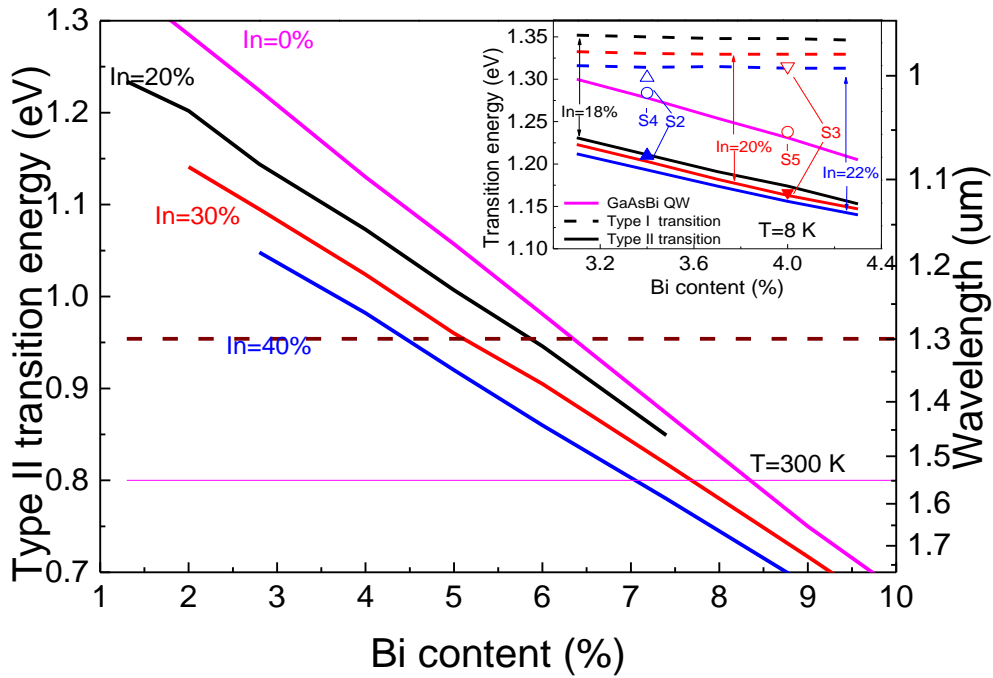

Figure 4

Figure 4. Calculated room temperature variation of the type II transition energy as a function of Bi and In composition for of 8.5 nm InGaAs/7.8 nm GaAsBi/8.5 nm InGaAs quantum wells epitaxially grown on GaAs based on 8 bands $k \cdot p$ model. Also, the calculated transition energy of type I and II at 8 K is plotted in the inset where the experiment data obtain from PL under low excitation power of 20 mW are shown.

As shown in the inset of Fig. 4, for the InGaAs/GaAsBi QW with In content of 18%, 20% and 22%, the calculated type I transition energies at 8 K, are about 1.356, 1.336 and 1.315 eV, respectively, while it is hardly influenced by the Bi content varying from 3.4% to 4.4%. For the In content between 18% and 22%, the reduction coefficient of type II transition, are about 6 meV/In% and 63 meV/Bi% which is mainly contributed by the valence band edge raising up due to Bi incorporated in GaAs. For the experiment data as shown in the inset of Fig. 4, the type I transition energy of S2 and S3 observed in low temperature PL is near 1.310 eV, which is lower than the expected value of 1.336 eV for the In content to be 20%. However it could be understood when considering the possible formation of $In_{0.2}Ga_{0.8}AsBi$ near the InGaAs/GaAsBi interface as mentioned above. The type II transition energies obtain from low temperature (8 K) PL spectra with excitation of 20 mW of these two samples are very close to the calculated results for the strained $In_{0.2}Ga_{0.8}As$/GaAsBi W QWs with Bi content of 3.4% and 4.0% respectively.

## 4. Conclusion

In summary, InGaAs/GaAsBi heterostructures with a varying Bi content grown by gas source MBE were investigated and optical luminescence properties of the grown structures were studied. The structure characterizations of HRXRD and TEM show good crystalline. These structures display strong low temperature type II luminescence, the energy of which varies with the Bi content and ranges from 1.16 to 1.20 eV with the FWHM typically 140 meV at 8 K. The transition energy of the W InGaAs/GaAsBi/InGaAs QWs and the single 7.8 nm GaAsBi/GaAs QWs are calculated basing on 8 bands $\mathbf{k \cdot p}$ model. The calculated values are very close to the transition energies obtain from PL. From the above discussion, we can see that using InGaAs/GaAsBi type-II structure is a promising way to extend the emission wavelength and realize GaAs-based diode lasers emitting near 1.3 μm.

## Acknowledgements

The authors wish to acknowledge the financial support of the National Basic Research Program of China (Grant No. 2014CB643902), the Key Program of Natural Science Foundation of China (Grant No. 61334004), the Natural Science Foundation of China (Grant No. 61404152), the "Strategic Priority Research Program" of the Chinese Academy of Sciences (Grant No. XDA5-1), the foundation of National Laboratory for Infrared Physics, the Key Research Program of the Chinese Academy of Sciences (Grant No. KGZD-EW-804) and the Creative Research Group Project of Natural Science Foundation of China (Grant No. 61321492).